\documentclass[aps,prl,reprint,superscriptaddress,10pt,showpacs,a4paper]{revtex4-1}
\newcommand{\papertitle}{Enstrophy Cascade in Decaying Two-Dimensional Quantum Turbulence}
\usepackage{amsmath,amssymb,amsfonts}
\usepackage{mathrsfs}
\usepackage{empheq}
\usepackage{graphicx}
\usepackage{txfonts}
\usepackage{color}
\usepackage{math tools}
\usepackage[unicode,breaklinks]{hyperref}
\usepackage[super]{nth}
\usepackage{braket}
\usepackage{tabularx}
\hypersetup{
    unicode=true,
    a4paper=true,
    plainpages=false,
    pdftitle={\papertitle},
    pdfauthor={M. T. Reeves, T. P. Billam, X. Yu, and A. S. Bradley},
    pdfsubject={\papertitle},
    colorlinks=true,
    linkcolor=blue,
    citecolor=blue,
    filecolor=black,
    urlcolor=blue
}
\DeclareSymbolFont{symbolstx}{OMS}{txsy}{m}{n}
\DeclareSymbolFontAlphabet{\mathcal}{symbolstx}

\newcommand{\eqnsrefp}[2]{{[Eqs.~(\ref{#1}) and (\ref{#2})]}}
\newcommand{\eqnreft}[1]{{Eq.~(\ref{#1})}}

\newcommand{\eqnsreft}[2]{{Eqs.~(\ref{#1}) and (\ref{#2})}}
\newcommand{\figreft}[2]{Fig.~\ref{#1}#2}

\newcommand{\xx}{\mathbf{x}}
\newcommand{\rr}{\mathbf{r}}
\newcommand{\vv}{\mathbf{v}}
\newcommand{\kk}{\mathbf{k}}
\newcommand{\ww}{\mathbf{w}}

\newcommand{\TT}{\tilde{T}}

\newcommand{\re}{\operatorname{\mathrm{R\kern-.045em e}}} 
\newcommand{\res}{\re_\mathrm{s}}

\newcolumntype{b}{X}
\newcolumntype{s}{>{\hsize=.5\hsize}X}

\newcommand{\otago}{Department of Physics, Centre for Quantum Science, and
Dodd-Walls Centre for Photonic and Quantum Technologies, University of Otago,
Dunedin, New Zealand} 

\newcommand{\queensland}{Australian Research Council Centre of Excellence in Future Low-Energy Electronics Technologies, School of Mathematics and Physics, University of Queensland, St Lucia, QLD 4072, Australia.}

\newcommand{\newcastle}{Joint Quantum Centre (JQC)
Durham--Newcastle, School of Mathematics and Statistics, \\ Newcastle
University, Newcastle upon Tyne, NE1 7RU, United Kingdom}

\begin{document}

\title{\papertitle}

\author{Matthew T. Reeves}
\email{m.reeves@uq.edu.au}
\affiliation{\otago}
\affiliation{\queensland}

\author{Thomas P. Billam}
\email{thomas.billam@newcastle.ac.uk}
\affiliation{\newcastle}

\author{Xiaoquan Yu}
\affiliation{\otago}

\author{Ashton S. Bradley}
\affiliation{\otago}

\date{\today}

\begin{abstract}
We report evidence for an enstrophy cascade in large-scale point-vortex
simulations of decaying two-dimensional quantum turbulence. Devising a method to generate
quantum vortex configurations with kinetic energy narrowly localized near a single length scale, the dynamics are found to be well-characterised by a superfluid
Reynolds number, $\res$, that depends only on the number of vortices and
the initial kinetic energy scale. Under free evolution the vortices exhibit 
features of a classical enstrophy cascade, including a $k^{-3}$
power-law kinetic energy spectrum, and steady enstrophy flux
associated with inertial transport to small scales. Clear signatures of the
cascade emerge for $N\gtrsim 500$ vortices. Simulating up to very large
Reynolds numbers ($N = 32, 768$ vortices), additional features of the
classical theory are observed: the Kraichnan-Batchelor constant  is found to
converge to $C' \approx 1.6$, and the width of the $k^{-3}$ range scales as
$\res^{1/2}$. The results support a universal phenomenology underpinning classical and quantum fluid turbulence.
\end{abstract}

\maketitle

Quantum vortices in atomic Bose-Einstein condensates (BECs)
offer the possibility not only to physically realize the point-vortex
model envisaged by Onsager \cite{Onsager1949}, but also to observe and
manipulate it at the level of individual quanta.  Experimental
techniques to controllably generate quantum vortices \cite{Kwon2014, Neely2010,
Wilson2013, Samson2016}, produce hard-wall trapping potentials containing
large, uniform density condensates \cite{Henderson2009, Gauthier2016}, and
determine vortex circulation \cite{Seo2016} have all been recently
demonstrated, and measurements of thermal friction coefficients \cite{Moon2015}
and vortex annihilation and number decay \cite{Kwon2014, Neely2013} have
already been made.  For well-separated vortices, the point-vortex regime of two-dimensional quantum
turbulence (2DQT) can be considered as a `stripped-down' model of hydrodynamic
turbulence with a definite number of degrees of freedom \cite{Billam2015}, and thus studying the analogies between 2QDT and
2D classical turbulence (2DCT) may expand our understanding of universal 
turbulent phenomena.  The recent experimental
observation of a von K\'arm\'an vortex street and the transition to turbulence in
the wake of a stirring obstacle~\cite{Kwon2016}, for example, adds to evidence that the classical Reynolds number concept may be generalized to quantum turbulence in frictionless superfluid flows~\cite{Frisch:1992a,finne2003,Reeves2015}.

The enstrophy cascade of decaying 2DCT predicted by
Batchelor \cite{Batchelor1969} is a key process of classical turbulence for which the quantum analogue has remained unexplored. While much theoretical attention has focused on
the inverse energy cascade of forced
turbulence~\cite{Numasato2009, Numasato2010, Reeves2013, Chesler2013a,
Kobyakov2014, Siggia1981} and macroscopic vortex clustering in
2DQT~\cite{Billam2014, Simula2014a,Yu2016}, a clear demonstration of an enstrophy
cascade has yet to be presented. A challenge to overcome in order to numerically
demonstrate such a cascade in 2DQT is that of obtaining sufficiently large
vortex number, initial spectral energy concentration, and range of wave numbers $k$, to identify the steep
associated energy spectrum, $E(k) \propto k^{-3}$, over a significant range of scale
space. The $k^{-3}$ scaling must also occur at large enough scales
to distinguish it from the identical, physically unrelated, power-law scaling in the kinetic energy spectrum at the vortex-core scale~\cite{Bradley2012a}. 

In this Letter we directly simulate an $N$-point-vortex model of decaying 2D quantum turbulence at large $N$. We devise a method of constructing an
initial condition with a large energy contained within a single wavenumber,
allowing us to simulate the 2DQT analog of a scenario where the existence of an
enstrophy cascade is well-established in 2DCT~\cite{Fox2010,Lindborg2010}. The
initial states are found to be well-characterised by a superfluid Reynolds
number $\res$ that depends only on the number of vortices and the initial
wavenumber $k_i$. We show that under free evolution the characteristic $k^{-3}$
spectrum of the enstrophy cascade emerges for $N \gtrsim 500$, and the associated
enstrophy and energy fluxes are found to agree with the Batchelor theory. By
increasing $N$ up to $32,768$, additional key features of the theory are
verified: the Kraichnan--Batchelor constant is found to be $C' \approx  1.6$,
close to the accepted classical value, and the length of the inertial range
scales as $\res^{1/2}$.

\emph{Background.---} Turbulent flows at large Reynolds numbers
($\re$) can spontaneously develop self-similar \emph{cascade} solutions, in
which quantities are conservatively transported across a subregion of scale
space called the inertial range. Two-dimensional turbulence cannot support the
usual Kolmogorov energy cascade of 3D turbulence, since the mean square
vorticity, or \emph{enstrophy} is unable to be
amplified through vortex stretching. However, Batchelor~\cite{Batchelor1969}
hypothesised that in 2D the enstrophy itself could therefore undergo a  cascade, from small to
large wavenumbers, via a filamentation of vorticity patches. The enstrophy
cascade is signified by a kinetic energy spectrum $E(k) =C' \eta^{2/3} k^{-3}$,
where $\eta$ is the enstrophy dissipation rate (assumed equal to the enstrophy flux in
the inertial range), and $C'$ is the Kraichnan--Batchelor constant. The
lossless cascade terminates at a dissipation wavenumber $k_d \sim k_i
\re^{1/2}$, at which viscous dissipation becomes important. The enstrophy
casade must be accompanied by a drift of energy to small wavenumbers, in order to
be simultaneously consistent with the conservation laws of energy and enstrophy. 

\emph{ Model.---} We consider a quantum fluid, such as a BEC, characterized by
healing length $\xi$ and speed of sound $c$, carrying quantized vortices of
charge $\kappa_i = \pm 1$ and circulation $\Gamma_i = \kappa_i \Gamma$
\footnote{In the case of an atomic BEC, one has $\xi = \hbar / \sqrt{\mu m}$,
$c = \sqrt{\mu/m}$, and $\Gamma = h/m$, where $\mu$ is the chemical potential
and $m$ is the mass of a constituent particle.}. For a quasi-2D system, vortex
bending is suppressed and the dynamics become effectively
two-dimensional~\cite{Rooney2011}.  In the low Mach number limit, where the
average intervortex distance $\ell$ is much greater than the healing length
$\xi$, interactions between vortices and density fluctuations 
can be ignored on
scales $\gtrsim \xi$. In this limit a fully compressible (e.g.,
Gross-Pitaevskii~\cite{Blakie2008}) description, that complicates
interpretation of fluxes \cite{Billam2015}, is not needed.  Instead, the motion
of the $i$th quantum vortex, located at $\rr_i$, can be described by a
dissipative point-vortex model \cite{tornkvist_shroder_prl_1997} with
compressible effects (at length scales $\lesssim \xi$) added phenomenologically
\cite{Billam2015,Kim2016}. The motion of the $i$th quantum vortex, located at
$\mathbf{r}_i$, is given by
\begin{align}
\frac{d \rr_i}{dt} &= \vv_i + \ww_i;\;\; \vv_i = \sum_{j=1,j\neq i}^N
\vv_{i}^{(j)};\;\; \ww_i = -\gamma \kappa_i \hat{\mathbf{e}}_3 \times \vv_{i},
\label{eqn:vortex_eom}
\end{align}
where $\gamma$ is  the dissipation rate, $\mathbf{\hat e}_3$ is a unit vector
perpendicular to the fluid plane, and $\vv_i$ and $\ww_i$ are the conservative
and dissipative parts of the velocity respectively. The
dissipation rate $\gamma$ arises from thermal friction due to the normal fluid
component, here assumed to be stationary \cite{Moon2015}.
Phenomenologically, we remove opposite-sign vortex
pairs separated by less than $\xi$ (modelling dipole annihilation), and smoothly increase the dissipation
$\gamma$ for same-sign vortex pairs as their separation decreases to around
$\xi$ (modelling sound radiation by accelerating vortices~\cite{pismen}). Details can be found in the Supplemental Material~\cite{SuppInfo}, or Ref.~\cite{Billam2015}.

The velocity of the $i$th vortex due to the $j$th,  $\vv_{i}^{(j)}$, is
obtained from a Hamiltonian point-vortex model subject to appropriate boundary
conditions.  As usually considered classically
\cite{Lilly1969,Lilly1971,brachet1988,Vallgren2011},  we will consider a
doubly-periodic square box with side length $L \gg \xi$, for
which~\cite{weiss_mcwilliams_pf_1991}
\begin{equation}
\vv_{i}^{(j)} = 
  \frac{\pi c \kappa_j } {(L/\xi)}
  \sum_{m=-\infty}^{\infty}
  \left( \begin{array}{c}
    \frac{-\sin(y_{ij}^\prime)}{\cosh(x_{ij}^\prime-2\pi m) - \cos(y_{ij}^\prime)}\\
    \frac{\sin(x_{ij}^\prime)}{\cosh(y_{ij}^\prime-2\pi m) - \cos(x_{ij}^\prime)}
  \end{array} \right),\label{eqn:periodic_v}
\end{equation}
where $(x_{ij}^\prime,y_{ij}^\prime)/(2\pi/L) \equiv \rr_{ij} \equiv \rr_i -
\rr_j$.  The absence of a physical boundary offers the usual advantage:
vortices cannot reach their own images, enforcing conservation of the (zero)
net vorticity. This helps achieve statistical homogeneity and isotropy, as
required for comparisons with Batchelor's theory.

\emph{Spectrum.---} The kinetic energy spectrum (per unit mass) in the periodic
box is given by~\cite{Billam2014} 
\begin{align}
E(\kk) ={}& E_{\rm self}(\kk) + E_{\rm int}(\kk)  \\
={}&  \frac{\Gamma^2}{8( \pi k
L)^2}\left[ N +  2\sum_{i=1}^N \sum_{j=i+1}^{N} \langle \kappa_i \kappa_j
\cos(\mathbf{k}\cdot\mathbf{r}_{ij}) \rangle \right],\label{eqn:spectrum}
\end{align}
where $\kk = (n_x \Delta k, n_y \Delta k)$ for  $n_x,n_y \in \mathbb{Z}$,
$\Delta k = 2\pi/L$, and $\langle \,\cdot\, \rangle$ denotes ensemble
averaging. The average kinetic energy is $\sum_\kk \; E(\kk) (\Delta k)^2 =
E_{\rm self} + E_{\rm int}$.  
The self-energy term is, for fixed $N$, a
cutoff-dependent constant, set by $L$ and the vortex core structure at wavenumbers $k \gtrsim \xi^{-1}$~\cite{Billam2014} (not considered here).
The time evolution of $E(\kk)$
governs the spectral transport of kinetic energy:
\begin{align}
\frac{dE(\kk)}{dt} &= T(\kk) + D(\kk),
\end{align}
where $T(\kk)$ is the transfer function, given by
\begin{align}
T(\kk) &= -\frac{\Gamma^2}{4 (\pi k L)^2}\sum_{i=1}^N \sum_{j=i+1}^N \langle \kappa_i
\kappa_j \sin(\mathbf{k}\cdot\mathbf{r}_{ij}) \kk \cdot ( \vv_i - \vv_j ) \rangle, \label{eqn:tk} 
\end{align}
and $D(\kk)$ is the dissipation spectrum, obtained from \eqnreft{eqn:tk} by
setting $\vv \rightarrow \ww$.  As usual, the enstrophy and energy spectra are
related via $\Omega(\kk) = 2k^2 E(\kk)$. Like its classical counterpart, the
superfluid transfer function $T(\kk)$ conservatively redistributes energy, with
$\sum_\kk T(\kk) (\Delta k)^2= 0$. The dissipation spectrum $D(\kk)$ governs the rate of energy loss: $\sum_\kk D(\kk) (\Delta k)^2 = dE/dt < 0$.
The one-dimensional (angularly integrated) spectral measures $E(k) = \int  d\phi_k k E(\kk)$ etc., are analysed by defining a discrete angular integral over a ring of wavenumbers: $ \tilde{f} (n \Delta k) = \sum_{\kk \in \mathcal{D}_n } f(\kk)
\,\Delta k$, where $\mathcal{D}_n = \{ \kk \, | \, (n-1/2)\Delta k \leq
|\mathbf{k}| \leq (n+1/2)\Delta k \}$, and $n=1,2,\ldots$. Hence we may define the discrete energy
and enstrophy fluxes~\cite{Kraichnan1967,Kraichnan1980}
\begin{align}
\tilde{\Pi}_\epsilon \left( n\Delta k \right)  &= - \textstyle \sum_{m=1}^n \TT(m\Delta k)\, \Delta k,\label{eqn:boxflux} \\
\tilde{\Pi}_\omega \left( n\Delta k \right) & = - 2\textstyle \sum_{m=1}^n (m \Delta k)^2 \TT(m\Delta k)\, \Delta k, \label{eqn:boxflux2}
\end{align}
that represent the instantaneous energy and enstrophy fluxes through the
$k$-space bin $|\kk| = n\Delta k$ due to the conservative
interactions. Turbulent cascades can be expected to develop when $\gamma \ll 1$ and
$T(\kk)$ is large, allowing a lossless inertial range to be
established over some range of $k$.

\begin{figure}
\includegraphics[width = 0.89\columnwidth]{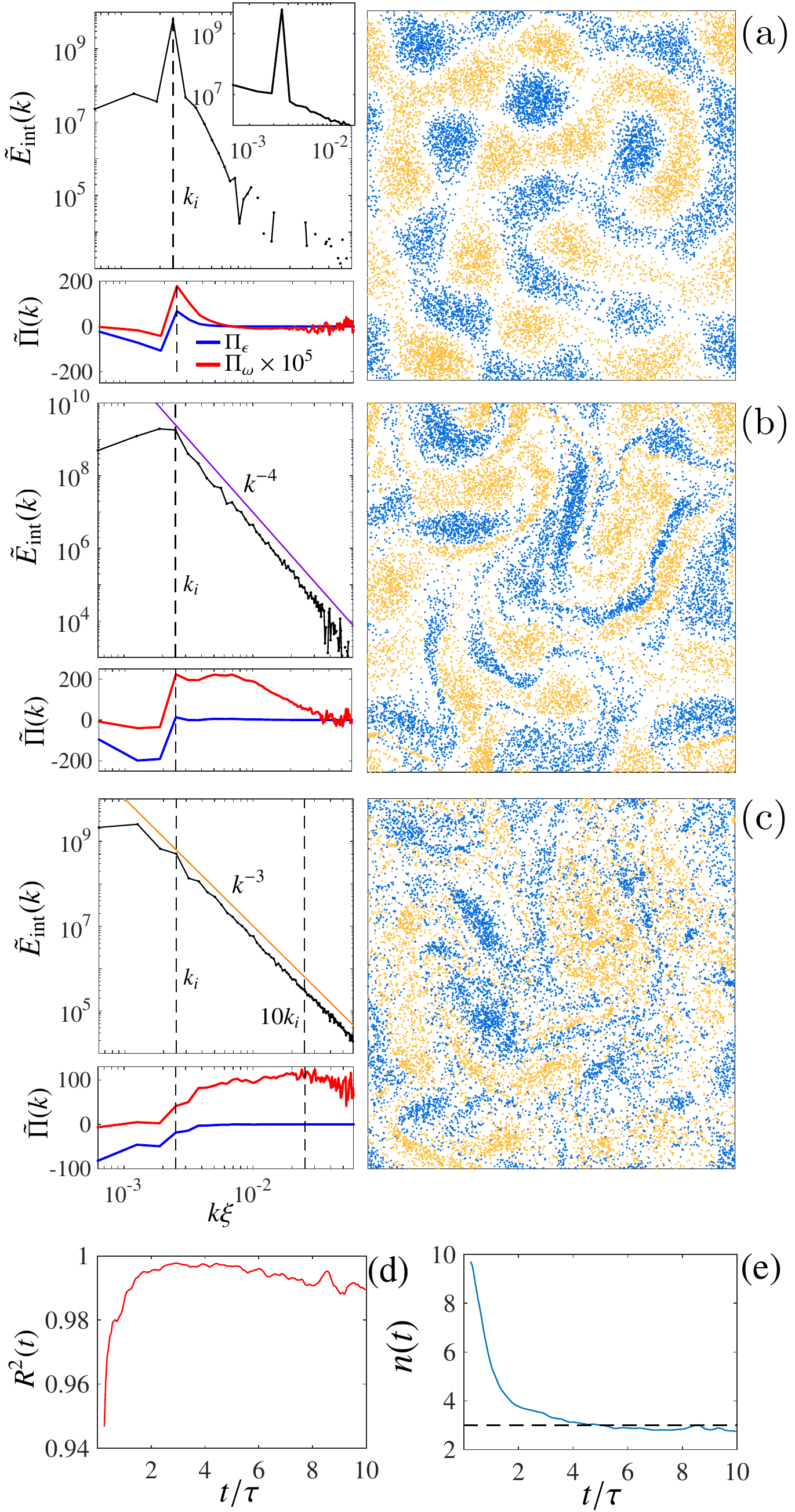}
\caption{ (color online). Vorticity distributions, kinetic energy spectra, and
fluxes (in units of $\Gamma^2/4\pi
L^2$) for $N = 16,384$ at (a) $t\approx 0.25\tau$, (b) $t \approx 1.85\tau$,
and (c) $t\approx 4\tau$. In the top panel, at high $k$, $E_{\rm int}(k)$ is
comparatively small and oscillates about zero. The negative values cannot be
shown on the log scale, causing the broken data line. Inset shows the full
spectrum, $E_{\rm self}(k) + E_{\rm int}(k)$, at $t=0$. (d) and (e) show, respectively, the
$R^2$ goodness of fit value and the best-fit slope $n$ [$E(k) \sim k^{-n}$] as functions of time, obtained from a linear fit
to the log-log data over the decade of wavenumbers marked in (c).
Values are averaged over $4$ runs and $9$ time samples with time-spacing $\delta
t \approx\tau/20$. The dashed line in (e) shows $n=3$.}
\label{fig:spectra}
\end{figure}

\emph{Initial condition.---} Although at sufficiently large Reynolds number
any initial state should tend towards the $k^{-3}$ similarity state, the simplest initial state has all the kinetic energy localized
near an initial wavenumber, as is often considered classically
\cite{Lilly1971,Lindborg2010}. However, it is not immediately evident from
\eqnreft{eqn:spectrum} how such a state can be created with quantum vortices.
Here we devise a simple method to create a superfluid analog of these states:
We define a set of wavenumbers $\mathcal{D}_i$ that form a shell of width $w$
localised around a chosen initial wavenumber $k_i$: $\mathcal{D}_i= \{ \kk \, |
\, k_i - w/2 \leq |\mathbf{k}| \leq k_i + w/2\}$.  Each mode in $\mathcal{D}_i$
is occupied with a random complex phase $\theta(\kk)$, uniformly sampled on
$[0,2\pi]$ to define a (Hermitian) vorticity field $\hat \omega(\kk) =
e^{i\theta(\kk)}$ if $\kk \in \mathcal{D}_i$ and $\hat \omega(\kk) = 0$ otherwise. The
real-space vorticity field, $\omega(\rr) = \int d^2\kk\; e^{i\kk\cdot \rr} \hat
\omega(\kk)$,  is then separated into positive and negative regions
$\omega_\pm(\rr)$ as  $\omega_+(\rr) = \omega(\rr)$ if $\omega(\rr) >0$ and
$\omega_+(\rr) = 0$ otherwise, and similarly for $\omega_-(\rr)$.  The
components are then normalised to unity [$\int d^2\rr \omega_\pm(\rr) = 1$],
and used as probability distributions to create an $N$-point-vortex initial
condition via rejection sampling.  This procedure creates an initial condition
with the vast majority of the interaction energy contained within one $k$-mode
[Fig.~\ref{fig:spectra}(a), inset], even for small vortex numbers $N\sim
10^2$. 

\emph{Dynamics.---} Starting from the initial conditions described above, we
simulate the dynamics of neutral point-vortex systems with fixed $L= 10^4\xi$,
fixed dissipation $\gamma = 10^{-4}$~\cite{Bradley2012a,Stagg2014a}, and vortex
numbers $N = 2^n, n = \{9,10,\dots,15\}$ \footnote{The point-vortex
approximation requires that $u_{\rm rms} \ll c$, so in this sense, for the
given parameters, the largest $N$ simulations are not physically reasonable.
However, our choice of $L$ is somewhat arbitrary, and the rescaling $\{\xx,L\}
\rightarrow \lambda \{\xx,L\}$ (for fixed $\xi$) yields  $\mathrm{v}_{\rm rms} \rightarrow \mathrm{v}_{\rm
rms}/\lambda$.}.  The
system can be characterised by the superfluid Reynolds number $\mathrm{Re_s}$
\cite{Onsager1953} \footnote{The superfluid Reynolds number introduced in
Ref.~\cite{Reeves2015} may be more appropriate in the presence of a stirring potential. } and the eddy turnover
time~$\tau$
\begin{align}
\re_\mathrm{s} &= \frac{E_{\rm int }^{1/2}L_i}{\Gamma}, \quad \quad \tau = \frac{L_i}{\mathrm{v}_{\rm rms} },
\label{eqn:res_and_tau}
\end{align}
where $L_i = 2\pi/k_i$, and $\mathrm{v}_{\rm rms}$ is the root-mean-square vortex velocity.
In the Supplemental Material~\cite{SuppInfo} we show that  for a wide range of the localised
initial conditions, $E_{\rm int}$ is well-approximated by
\begin{align}
E_{\rm int} &= A \times \left( \frac{\Gamma^2}{4\pi L^2} \right) \left(\frac{N}{k_i}  \right)^2 (\Delta k)^2,
\end{align}
where $A = \mathrm{const.} \approx 0.25$. Neglecting unimportant
constant factors, this yields a remarkably simple formula for $\mathrm{Re_s}$
as the ratio of two dimensionless quantities,
\begin{align}
\mathrm{Re_s} &= N/n_i^2,
\label{eqn:res}
\end{align}
where $n_i \equiv k_i/\Delta k$ is the dimensionless initial wavenumber.  To
maximise $\mathrm{Re_s}$ while still maintaining approximate isotropy, we thus
set $L_i = L/4$. Since $\mathrm{Re_s}$ is independent of the value of $w$, we
choose the narrowest window, $w=\Delta k $. We directly simulate the
point-vortex model \eqnsrefp{eqn:vortex_eom}{eqn:periodic_v} and compute time-
and ensemble-averaged spectra and fluxes
[Eqs.~(\ref{eqn:spectrum})--(\ref{eqn:boxflux2})], using GPU codes
\cite{cudaRef} that allow us to evaluate the full $N$-body problem for very
large $N$.

Fig.~\ref{fig:spectra}(a)-(c) shows the dynamics of the vortices, kinetic
energy spectra, and fluxes for $N=16,384$. The qualitative behaviour is similar
for all $N$ considered, but naturally large $N$ yields cleaner results.
Movies for some cases are provided in the Supplemental
Material~\cite{SuppInfo}. Very early times
[Fig.~\ref{fig:spectra}(a)] show the spectrum rapidly spreads from the initial
state well-localised at $k_i = 4(\Delta k)$ [Fig.~\ref{fig:spectra}(a), inset].
A linear fit to the log-log spectrum indicates that $t \approx 2 \tau$,  where the $R^2$ goodness of fit plateaus near unity [Fig~
\ref{fig:spectra}(d)], marks  the onset of power-law scaling.  At the onset,
the spectrum agrees quite well with the Saffman~\cite{saffman1992} scaling $k^{-4}$, consistent with the formation of sharp, isolated
vorticity-gradient filaments [Fig.~\ref{fig:spectra}, (b)].  These filaments
are repeatedly stretched and packed, and the spectral slope gradually
transitions, settling to the $k^{-3}$ scaling from $t\sim 4 \tau$ onwards
[Fig.~\ref{fig:spectra}, (c,e)], maintaining a high goodness of fit, $R^2 >0.988$ [Fig~\ref{fig:spectra}(d)]. A transition from $k^{-4}$ to $k^{-3}$ scaling
was also reported in pseudospectral Navier-Stokes simulations of decaying 2D
turbulence~\cite{brachet1988}. Note that in Fig.~\ref{fig:spectra} only the
interaction term $E_{\rm int }(k)$ is shown, as the self-energy term can only
ever contribute a trivial $N/k$ scaling. 

Inspection of the energy and enstrophy fluxes confirms the directions of
spectral transport. The early developing stages of evolution
[Fig.~\ref{fig:spectra}(b)] clearly demonstrate a development of a negative
energy flux (indicating flow to low $k$) and positive enstrophy flux (indicating
flow to high $k$) in the mutually exclusive wavenumber regions $k < k_i$ and $k
> k_i$ respectively.  The $k^{-3}$ spectrum [Fig.~\ref{fig:spectra}(c)] is
corroborated by a nearly constant enstrophy flux over approximately one decade
of wavenumbers, providing a means to estimate $\eta$ and determine the
Kraichnan-Batchelor constant via the so-called compensated kinetic energy
spectrum: $C' = E(k) k^3/\eta^{2/3}$, where $\eta = \Pi_\omega$ averaged over
$k$, time window, and ensemble. 

\begin{figure}[t] \includegraphics[width = \columnwidth]{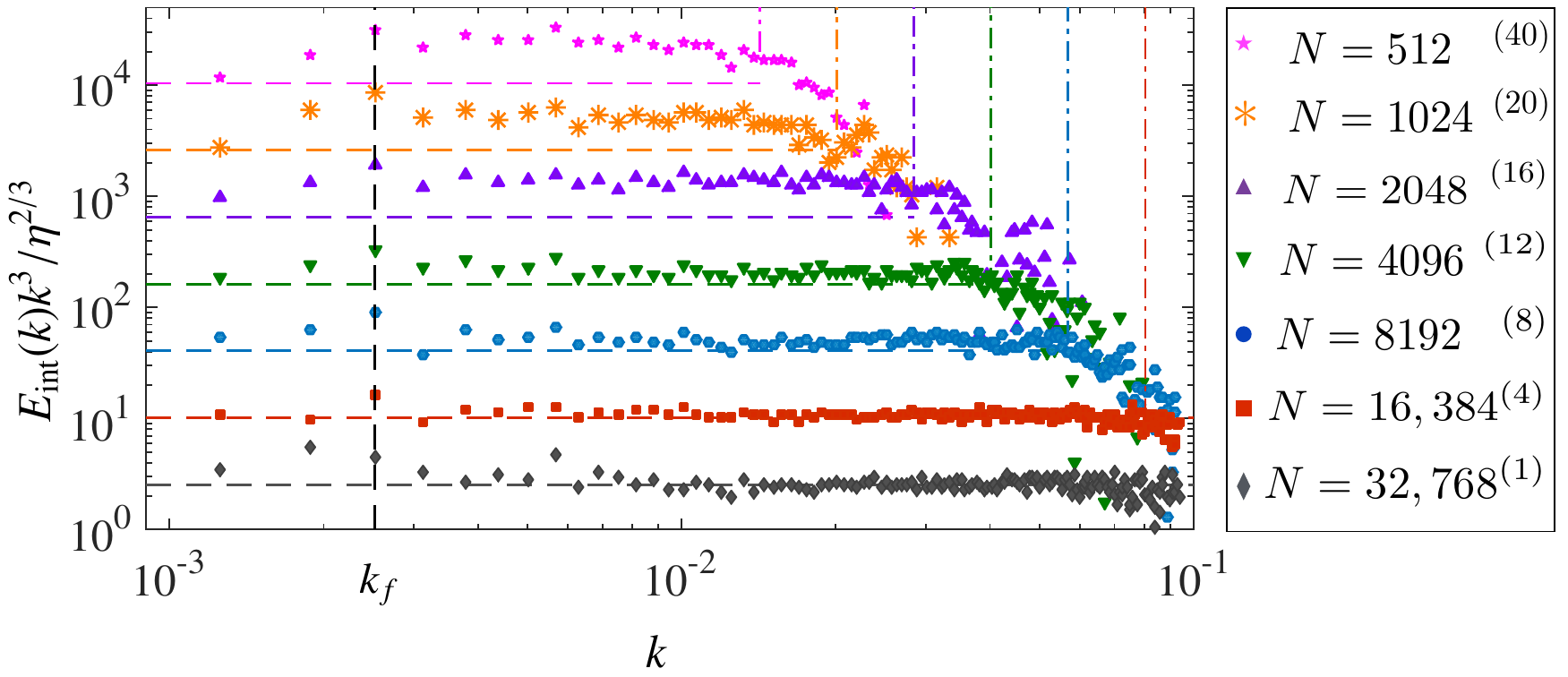} 
\caption{ Compensated
kinetic energy spectra for a range of $N$, averaged over ensemble and a time
window $\sim0.5\tau$. For clarity the spectra are vertically shifted by
increasing powers of $4$. The horizontal dashed lines shows the value $C' =
1.6$ (also vertically shifted for comparison at different $N$). Dash-dot lines
indicate the intervortex distance wavenumber $k_\ell$ at different $N$ (see
text).  In the legend the bracketed superscripts indicate the number of
independent realisations used in the ensemble average. }
\label{fig:KraichnanBatchelor}
\label{spectraVsN} \end{figure}

\figreft{fig:KraichnanBatchelor}~shows the compensated spectrum for different
$N$ at $t \sim 4 \tau$. The $k^{-3}$ scaling is observed to some degree for all
$N$ considered, albeit over less than a decade for small $N$ ($\sim0.7$ decades
for $N=512$). However the quality and range of the scaling increases
dramatically as $N$ is increased. For smaller $N$, $C'$ is quite large ($C'
\approx 3.8$) \footnote{Slight variation of $C'$ with the forcing scale or
Reynolds number is not uncommon~\cite{Kraichnan1967, Lindborg2010,
Vallgren2011}}, but as $N$ increases $C'$ decreases and tends  towards a
constant value $C' \approx 1.6$. A simulation with $N=16,384$ and $k_f = 8
(\Delta k)$ yielded $C' \approx 2.0$, in good agreement with $N=4096$, $k_f = 4
(\Delta k)$, that has the same $\res$ and yielded $C'\approx 1.9$. 
The scaling range is found to persist up to $k_\ell = 2\pi /\ell$, the
wavenumber associated with the average intervortex distance $\ell = L/N^{1/2}$.
Notice that for $N\geq16,384$ this means the compensated spectrum is constant
over a significant range of roughly 1.5 decades above the initial wavenumber.
Above $k_\ell$, the interaction spectrum quickly decreases, indicating a
transition from many-vortex to single-vortex physics. 

\emph{Discussion.---} It appears that the basic phenomenology of the decaying
enstrophy cascade can indeed be seen in 2DQT. For large $\res$, we find a
Kraichnan-Batchelor constant $C' \approx 1.6$ close to the accepted value for a
classical fluid, $C' =1.4$ \cite{Paret1999,Lindborg2010}.  Similarly, the Kolmogorov
constant in 3D has been found to be the same above and below the
$\lambda$-transition in superfluid He$^4$ \cite{barenghi2014}.  Our observation
of greater values of $C'$ at lower $\res$ (although with greater uncertainties) suggests 
that fewer available degrees of freedom result in less efficient spectral transport.
Importantly, our results 
show that $\res$ as defined in \eqnreft{eqn:res} quantifies
the degree of turbulence very well, and can be used to estimate the range of the enstrophy 
cascade.  Since $\res = N (\Delta k/k_i)^2$, our results exhibit the same power-law 
range scaling as a
classical fluid: $ k_\ell/k_i \sim \res^{1/2}$. Similarly, a recent experiment~\cite{Babuin2014} found  $\ell^{-1} \sim \res^{3/4}$ in 3DQT (for an appropriately defined $\res$), similar to the dissipation scale in the Kolmogorov energy casade. However, here the cascade
terminates due to a crossover from many-vortex to single-vortex physics, rather
than due to dissipative effects. The sudden drop in $E_{\rm int}(k)$ at
$k_\ell$ suggests that the point-vortex system can effectively be truncated at
wavenumbers $k\sim k_\ell$, as was qualitatively argued by
Kraichnan~\cite{kraichnan1975}.

Further study of how dual inverse-energy and direct-enstrophy cascades
\cite{Kraichnan1967, Leith1968} could manifest in \textit{forced} 2DQT is
certainly warranted.  A study of the inverse energy cascade using a forced
point-vortex model \cite{Siggia1981} found the Kraichnan-Kolmogorov constant to
be twice the accepted value. However in Ref. \cite{Siggia1981} forcing was
introduced by essentially reversing the sign of $\gamma$. Adapting our
rejection-sampling method to \textit{dynamically} introduce vorticity instead
could provide a more physical model, roughly corresponding to turbulence
generated by a stirring grid in a 2D quantum fluid
\cite{Skaugen2016b,sommeria1986,rutgers1998}.  Studying the forced case would
allow exploration of conditions under which both cascades coexist in the
point-vortex system, and of intermittency effects \cite{Paret1999}.
It will also be interesting to explore the relation between the enstrophy cascade
observed here and the anomalous scaling at non-thermal fixed points in
compressible decaying 2DQT \cite{Karl2016, Schole2012}.

 Finally, let us  discuss the prospect of observing the cascade in atomic
condensates. The main challenge would be creating a system large enough
relative to the healing length, $\xi$. Currently, experimental setups have produced pure, stable condensates of up to $N_a \sim 10^8$ atoms with atomic
number densities $n_0 \sim 10^{14}$ cm$^{-3}$, using
$^{23}$Na~\cite{Streed2006}. The $s$-wave scattering length $a_s \approx
2.8$nm~\cite{Streed2006}, gives $\xi = 1/\sqrt{4\pi n_0 a_s}
\approx 0.53 \mu\mathrm{m}$. Assuming such parameters in a uniform quasi-2D
system of volume $L^2 h$, with thickness $h \sim 6.6\xi$~\cite{Kwon2014}, gives
$L/\xi \approx 1000$. This would allow $N = 512$, since here
$\mathrm{v_{rms}}/c \approx 0.025$ and hence the system size could be reduced
to $L/\xi \sim 850 $ without invalidating the incompressibility assumption
$\mathrm{v}_{\mathrm{rms}}/c \lesssim 0.3$ [in \eqnsreft{eqn:vortex_eom}{eqn:periodic_v}
$\{\xx,L\} \rightarrow \lambda \{\xx,L\}$ gives $\vv \rightarrow \lambda^{-1}
\vv$, $t \rightarrow \lambda^2 t$]. However, a system with $L/\xi \sim 1000$
would correspond to a cloud $\sim$$500$ $\mu$m across, an order of magnitude
larger than in current experiments. Recent experimental 2DQT studies have
acheived $L \sim 500 \xi$  and $N\sim80$ vortices in harmonically confined
systems~\cite{Kwon2014,Moon2015,Kwon2015}, and hard-wall traps
\cite{Gaunt2013,Navon2016} with $L/\xi \gtrsim 200$~\cite{Gauthier2016}.
Ultracold Fermi gases, with a much shorter healing length, may also be a viable
alternative~\cite{Bulgac2016}. Condensate lifetimes $T> 60$ s are
common~\cite{Kwon2014}, giving $T \approx 4.6\times10^5$ $\xi/c$, or $T/\tau
\approx 460$, greatly exceeding the requirements here. Some further additional
freedom is possible by decreasing $\xi \propto a_s^{-1/2}$ through a Feshbach
resonance~\cite{Inouye1998,Chin2010}, although this would eventually enhance
three-body losses~\cite{Stenger1999}. Controlled stirring protocols show
promise for efficient cluster
injection~\cite{Reeves2015,Stagg2014a,Kwon2014}. While
challenging, the required experimental conditions are not inaccessible. 

\emph{Conclusion.---} We have numerically observed signatures of an enstrophy cascade in decaying 2DQT, including a $k^{-3}$
power-law spectrum, constant enstrophy flux over a wide inertial range, and a
Kraichnan-Batchelor constant converging to $C' \approx 1.6$ for large vortex
number.  We have shown that the extent of the inertial range scales as $\res^{1/2}$
for a suitably-defined superfluid Reynolds number, $\res$, that depends only on the number of vortices and the length scale where kinetic energy is initially concentrated.  The relevance of the classical cascade theory for describing decaying 2DQT
suggests an underlying universality of decaying turbulence phenomena. Signatures of the enstrophy cascade become observable for systems of a few
hundred vortices, and may soon be within reach of cold-atom 2DQT experiments.

We thank B. P. Anderson for many stimulating discussions and A. J. Groszek for valuable comments. A.S.B was supported by a Rutherford Discovery Fellowship administered by the Royal Society of New Zealand.

\section*{Supplemental Material}
\section{Point-vortex simulations }
As described in the main text, we simulate a weakly-dissipative point-vortex
model with added phenomenological treatment of the main effects arising from
compressibility of a quantum fluid when vortices approach each other at
healing-length scales. This model was described in Ref.~\cite{Billam2015}; we
summarize it here for convenience.

We consider a dissipative point-vortex model in which the motion of the $i$th
quantum vortex, located at
$\mathbf{r}_i$, is given by
\begin{align}
\frac{d \rr_i}{dt} &= \vv_i + \ww_i;\;\; \vv_i = \sum_{j=1,j\neq i}^N
\vv_{i}^{(j)};\;\; \ww_i = -\gamma \kappa_i \hat{\mathbf{e}}_3 \times \vv_{i},
\label{eqn:vortex_eom}
\end{align}
where $\gamma$ is the background dissipation rate, $\mathbf{\hat e}_3$ is a unit vector
perpendicular to the fluid plane, and $\vv_i$ and $\ww_i$ are the conservative
and dissipative parts of the velocity respectively [see Eq. (1) in the main text].
The added phenomenolgical treatment has two aspects:

(a) To model the annihilation of closely-spaced vortex -- antivortex dipoles,
at the end of every simulation timestep we remove any opposite-circulation
vortex pairs that are separated by distances less than the healing length
$\xi$.

(b) To model the effects of sound radiation by closely-spaced same-circulation
vortex pairs, for a vortex $i$ with nearest same-circulation neighbour $s$
located a distance $r_{is}$ away we compute a local disspation rate
\begin{equation}
\gamma_i = \mathrm{max}\left( \exp \left[ \ln(\gamma) \frac{r_{is}-r_1}{r_2-r_1} \right],\gamma \right)\,.
\end{equation}
When computing the evolution of vortex $i$, we replace the background
dissipation rate $\gamma$ with the local dissipation rate $\gamma_i$ in
\eqnreft{eqn:vortex_eom}.  We choose $r_2 = \xi$ and $r_1 = 0.1\xi$, although the results are insensitive to the precise values of these parameters. Increasing $r_2$ and $r_1$ by an order of magnitude did not qualitatively alter the results presented \cite{Billam2015}.

We emphasise that the inclusion of dissipation in our model is important to
describe the dynamics of typical experimental quantum fluids.  For example, in
the case of BEC experiments \eqnreft{eqn:vortex_eom} can be derived from the
damped Gross-Pitaevskii equation \cite{Billam2015},
\begin{equation}
i \frac{\partial \psi}{\partial t} = (1-i\gamma) \left[-\frac{1}{2} \nabla^2 + |\psi|^2 -1 \right] \psi \label{eqn:dgpe},
\end{equation}
(written here in dimensionless form), in the limit of large vortex separation.
\eqnreft{eqn:dgpe} can itself be derived, by neglecting noise terms, from a
rigorous microscopic treatment of a degenerate Bose-gas \cite{Blakie2008},
where the background dissipation rate $\gamma$ describes collisions between
condensate and non-condensate atoms.  \eqnreft{eqn:dgpe} has been shown to
provide a capable description of experimentally observable BEC dynamics, where
$\gamma$ (calculable \textit{a priori} from the microscopic treatment) is
typically of order $10^{-4}$ \cite{Bradley2013}. We also note that because we
consider systems with large average inter-vortex spacing, the rate of vortex --
antivortex annihilations modeled by phenomenological treatment (a) described
above is low; we find $\lesssim 1$\% of the original vortices are annihilated
during our simulations.  This emphasizes the fact that the spectral transport
of kinetic energy we observe is driven by the $N$-body vortex dynamics, rather
than by decay processes.

\section{Interaction energy and Reynolds Number }
\begin{figure}[!t]
\centering
\includegraphics[width=0.8\columnwidth]{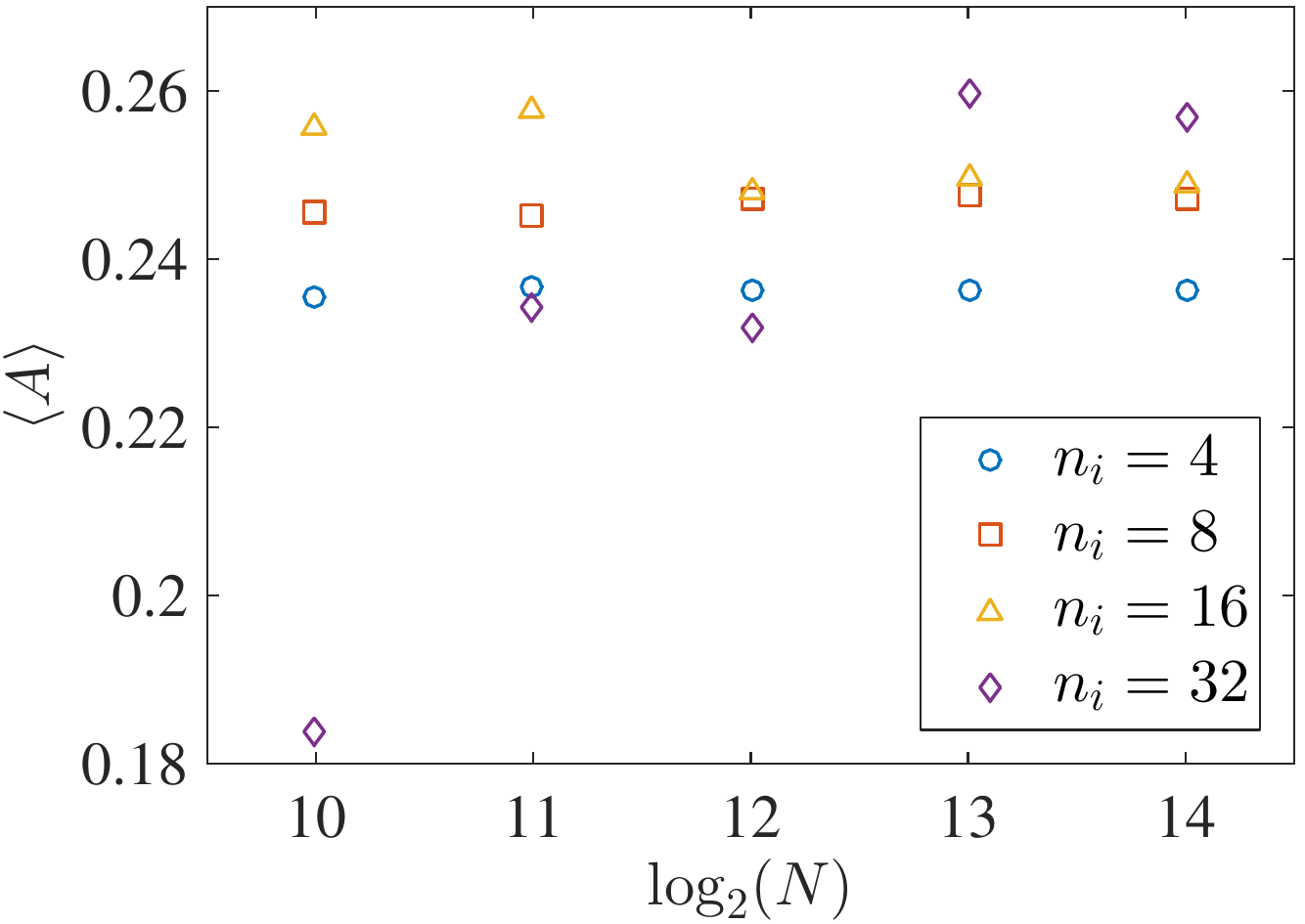}
 \caption{Average values for the parameter $\langle A \rangle$ as defined in \eqnreft{eqn:EintAnsatz}, for different values of the vortex number $N$ and the (dimensionless) initial wavenumber $n_i = k_i/\Delta k$. Averages were calculated from 100 samples for each value of $n_i$ and $N$.}
\label{fig:Eint_vs_N_and_nf}
\end{figure}

The point-vortex system can be characterized by Onsager's superfluid Reynolds number $\mathrm{Re}_s$
\cite{Onsager1953} and the characteristic eddy
turnover time~$\tau$
\begin{equation}
\mathrm{Re}_s = \frac{U D}{\Gamma}, \quad \quad \tau = \frac{D}{U},
\end{equation}
where $D$ and $U$ are an appropriate characteristic length and velocity respectively, and $\Gamma = h/m$ is the quantum of circulation. Natural choices to characterize $\tau$ are root-mean-square vortex velocity $U = \mathrm{v}_{\rm rms}$ and the initial cluster size $D = L_i = 2\pi/k_i$, thus defining a natural cluster turnover time. While $\mathrm{v_{rms}}$ could also be used for $\res$, it is equally valid to use $U = E_{\rm int}^{1/2}$, which has the same dimensions, and is a more natural parameter for characterizing the kinetic energy spectrum. This choice also allows for a useful formula for the Reynolds number to be obtained from the kinetic energy spectrum [Eq. (3) in the main text] 
\begin{multline}
E(\kk) = E_{\rm self}(\kk) + E_{\rm int}(\kk)  \\=  \frac{\Gamma^2}{8( \pi k
L)^2}\left[ N +  2\sum_{i=1}^N \sum_{j=i+1}^{N} \langle \kappa_i \kappa_j
\cos(\mathbf{k}\cdot\mathbf{r}_{ij}) \rangle \right],\label{eqn:spectrum}
\end{multline}
where $\kk = (n_x \Delta k, n_y \Delta k)$ for  $n_x,n_y \in \mathbb{Z}$,
$\Delta k = 2\pi/L$, and $\langle \,\cdot\, \rangle$ denotes ensemble
averaging. For the states with positive interaction energies of relevance here, the $N^2-N \approx N^2$ terms in the double sum of \eqnreft{eqn:spectrum} yield $E_{\rm int} \propto N^2$  (whereas at negative interaction energies $E_{\rm int} \sim N$, see, e.g.,~\cite{Montgomery1974a,Yu2016}). Furthermore, since the sum has been explicitly constructed to form  a
delta function shell of the radial wavevector, we are motivated to propose the (continuum) ansatz
\begin{equation}
\lim_{\Delta k \rightarrow 0}E_{\rm int}(\kk) = \frac{\Gamma^2}{8 ( \pi n \Delta k L)^2} \left[ N^2\frac{ \delta(n-n_i)}{n} \langle A(N,n_i) \rangle \right]
\end{equation}
where $n \equiv k/\Delta k$ and $n_i \equiv k_i/\Delta k \equiv L/L_i$ are dimensionless wavenumbers, and $A(N,n_i)$ is a random function that
allows for additional, ``anomalous" dependence on $N$ and $n_i$. For the continuum, making  the replacement  $\sum_\kk (\Delta k)^2 \rightarrow \iint n\, dn\, d\theta_n\, (\Delta k)^2$, yields
\begin{equation}
E_{\rm int} = \left( \frac{\Gamma^2}{4\pi L^2} \right) \left( \frac{N^2}{n_i^2} \right)  \langle A(N,n_i) \rangle.
\label{eqn:EintAnsatz}
\end{equation}
 The average values
$\langle A \rangle$  for a range of
$N$ and $n_i$, as calculated from the numerical initial conditions, are presented in Fig.~\ref{fig:Eint_vs_N_and_nf}. The value  $\langle A
\rangle$ is found to be virtually constant, and of order unity. The surprising result that $\langle A \rangle $ is close to constant leads to a remarkably simple formula for the Reynolds number as the ratio of two dimensionless quantities
\begin{equation}
\mathrm{Re_s} ' = \frac{N}{n_i^2}
\label{eqn:Res_ansatz}
\end{equation}
since we may formally neglect the factor of $\sqrt{\langle A \rangle/4\pi}$ when $\res'$ is large. \eqnreft{eqn:Res_ansatz} can be viewed as the product of the total vortex number and the typical cluster area (relative to the box area), which is essentially a measure of the number of vortices contained in each cluster. $\res'$ could therefore be interpreted as an \emph{effective} number of degrees of freedom, based on how important many-body effects are in the system due to same-sign vortex clustering.  The discrete vorticity field becomes uncorrelated when $n_i \gtrsim L/\ell$, where $\ell = L/\sqrt{N}$ is the average intervortex distance, since the discrete vorticity field will not be able to (on average) resolve spatial frequencies higher than $k \sim\ell^{-1}$. Hence by this measure, uncorrelated vortex distributions (i.e. the so-called ``ultraquantum" regime $E_{\rm int} \sim 0$) correspond to $\res' \sim 1$. One would expect the ansatz to become invalid as this regime is approached. Indeed this is clearly demonstrated by the deviation in the general trend in Fig.~\ref{fig:Eint_vs_N_and_nf} for the case $N=1024$, $n_i = 32$, for which $\res' = 1$. The requirement for \eqnreft{eqn:Res_ansatz} to be valid is therefore $n_i^2 \ll N$. States with negative interaction energies cannot be described by \eqnreft{eqn:Res_ansatz}.

\end{document}